\documentclass[aps,prl,twocolumn,showpacs,superscriptaddress,preprintnumbers,amsmath,amssymb,floatfix,10pt]{revtex4}
\usepackage{times}
\usepackage{graphicx}
\usepackage{dcolumn}
\usepackage{bm}
\usepackage{subfigure}
\usepackage{mathrsfs}
\usepackage{bbm}
\usepackage{textcomp}

\newcommand{\vect}[1]{\bm{#1}}

\renewcommand{\d}{\textnormal{d}}
\begin{document}
\title{Determining the carrier-envelope phase of intense few-cycle laser pulses}

\author{F. Mackenroth}

\author{A. Di Piazza}
\email{dipiazza@mpi-hd.mpg.de}

\author{C. H. Keitel}
\affiliation{Max-Planck-Institut f\"ur Kernphysik, Saupfercheckweg 1, 69117 Heidelberg, Germany}

\date{\today}
\begin{abstract}
The electromagnetic radiation emitted by an ultra-relativistic accelerated electron is extremely sensitive to the precise shape of the field driving the electron. We show that the angular distribution of the photons emitted by an electron via multiphoton Compton scattering off an intense ($I>10^{20}\;\text{W/cm$^2$}$), few-cycle laser pulse provides a direct way of determining the carrier-envelope phase of the driving laser field. Our calculations take into account exactly the laser field, include relativistic and quantum effects and are in principle applicable to presently available and future foreseen ultra-strong laser facilities.
\end{abstract}

\pacs{12.20.Ds,42.65.Re}
\maketitle

The rapid development of ultra-fast optics has been allowing the investigation of physical processes at shorter and shorter time scales. Time compression of laser pulses to durations $\tau$ below two laser cycles and even down to one cycle has been demonstrated in the mid infrared \cite{Bonvalet} ($\tau \approx 39\;\text{fs}$ at a central laser wavelength of $\lambda=12\;\text{$\mu$m}$), in the near infrared \cite{Krauss_2010} ($\tau \approx 4.3\;\text{fs}$ at $\lambda=1.55\;\text{$\mu$m}$), in the optical \cite{Cavalieri_2007} ($\tau < 4\;\text{fs}$ at $\lambda=0.7\;\text{$\mu$m}$) and in the extreme ultraviolet (XUV) domain \cite{Goulielmakis_2008} ($\tau \approx 80\;\text{as}$ at $\lambda=12\;\text{nm}$). Laser-matter interaction in this ``few-cycle'' regime shows features, which are qualitatively new with respect to the more conventional ``many-cycle'' regime: the response, for example, of atoms and molecules becomes sensitive to the precise temporal form of the electromagnetic field of the laser and, in particular, to its carrier-envelope phase (CEP), i. e. the phase difference between the carrier wave and the envelope function (see the recent review \cite{Krausz_2009} and the references therein). Viceversa, the knowledge and the control of the CEP of a laser pulse allows in turn to control physical processes like, for example, atomic ionization or above-threshold ionization (ATI). So far, few-cycle pulses have been produced with intensities below the relativistic threshold, corresponding in the optical domain to laser intensities $I$ of the order of $10^{18}\;\text{W/cm$^2$}$. Experimental determination of the CEP for few-cycle pulses of intensities up to $I = 10^{14}\text{-}10^{15}\;\text{W/cm$^2$}$ has been achieved by a stereo ATI measurement technique proposed in \cite{Paulus_2003}, with an accuracy of about $\pi/300$ \cite{Wittmann_2009}. Other methods employed to measure the CEP are attosecond streaking \cite{Goulielmakis_2004} and THz-spectroscopy \cite{Kress_2006}. However, these methods are not applicable for laser pulses of intensities above $I = 10^{16}$ W$/$cm$^{2}$ when relativistic effects become increasingly important.

The generation of intense laser pulses is intimately connected with temporal compression. Not only because tighter temporal compression implies, of course, larger pulse intensities at a given laser energy and waist size, but also because available larger intensities allow potential discovery of new physical processes at shorter time scales, opening the possibility of exploiting them to generate even shorter pulses. In the present context a laser pulse characterized by a peak electric field $\mathcal{E}$ and by a carrier angular frequency $\omega$, is indicated as  ``intense'', if the parameter $\xi=|e|\mathcal{E}/m\omega$ is much larger than unity, where $e<0$ is the electron's charge, $m$ its mass and units with $\hbar=c=1$ are employed, as throughout this work. An electron in such an intense laser field is ultra-relativistic, i. e. its Lorentz factor is much larger than unity \cite{Landau_b_2_1975}. In \cite{Tsung_2002} the possibility of generating single-cycle optical laser pulses with peak intensities larger than $10^{20}\;\text{W/cm$^2$}$, corresponding to $\xi> 10$, is investigated theoretically. Also, in \cite{Tsakiris_2006} the production of few-cycle, intense XUV bursts is envisaged by employing relativistic harmonic generation by a planar target. Moreover, the Petawatt Field Synthesizer (PFS) laser system under construction in Garching (Germany) aims at optical laser intensities of the order of $10^{22}\;\text{W/cm$^2$}$ ($\xi\approx 100$) by compressing an energy of $5\;\text{J}$ to only $5\;\text{fs}$, corresponding to less than two laser cycles \cite{PFS}. Finally, at the Extreme Light Infrastructure (ELI) \cite{ELI} facility, unprecedented laser intensities of the order of $10^{25}\text{-}10^{26}\;\text{W/cm$^2$}$ are envisaged, with pulse durations of about $10\;\text{fs}$. Therefore it is highly desirable to have a procedure to determine the CEP of few-cycle laser pulses also when their intensity largely exceeds the relativistic threshold.

In the present Letter we provide a method of determining in principle the CEP of an intense ($I>10^{20}\;\text{W/cm$^2$}$) few-cycle laser pulse. The method exploits precisely the specific features of the electromagnetic spectrum emitted by an ultra-relativistic electron, in particular that it emits radiation almost exclusively in a narrow cone of aperture  $m/\epsilon_e\ll 1$ along its instantaneous velocity, with $\epsilon_e$ being the electron's energy at the emission time \cite{Landau_b_2_1975}. Since intense laser pulses are of interest here, the laser field is taken into account exactly. Moreover, at such high laser intensities and electron energies, quantum effects may play in general a crucial role and they are also included by performing the calculations in the framework of quantum electrodynamics (QED) in the Furry picture \cite{Landau_b_4}. This requires the exact solutions of the Dirac equation in the presence of the external field as initial and final electron quantum states. Therefore, we restrict the quantum case to scenarios where the external field can be approximated by a plane wave and employ electron Volkov states \cite{Landau_b_4}. If quantum effects are negligible our method can be generalized to hold in principle for an arbitrary external field and we will consider the case of a focused Gaussian beam. We show that the angular distribution of the radiation emitted by the electron is particularly sensitive to the CEP of the driving pulse and that a theoretical accuracy of about $\pi/10$ can in principle be achieved. We also derive an analytical formula connecting the angular aperture of the spectrum with the value of the CEP.

We start by describing the few-cycle laser beam as a pulsed plane wave linearly polarized along the $x$ direction and propagating along the positive $z$ direction. The amplitude and the central angular frequency of the pulse are indicated as $\mathcal{E}$ and $\omega$, respectively. Then, the electric field $\vect{\mathcal{E}}(\phi)$ of the wave depends only on the phase $\phi=k^{\mu}x_{\mu}=\omega(t-z)$, with $k^{\mu}=\omega(1,0,0,1)$ and can be written as $\vect{\mathcal{E}}(\phi)=\mathcal{E}\psi_{\mathcal{E}}(\phi)\,\hat{\vect{x}}$, with $\psi_{\mathcal{E}}(\phi)$ being an adimensional function. In order to avoid the appearance of unphysical static (dc) components in the electric field \cite{Brabec_2000}, we choose the electromagnetic vector potential as $\vect{\mathcal{A}}(\phi)=\mathcal{A}\psi_{\mathcal{A}}(\phi)\,\hat{\vect{x}}$, where $\mathcal{A}=-\mathcal{E}/\omega$,
\begin{equation}
\psi_{\mathcal{A}}(\phi) = 
\begin{cases} \sin^4\left(\frac{\phi}{2N}\right)\sin(\phi+\phi_0) &  \phi\in[0,2N\pi]\\
0 & \text{elsewhere,}
\end{cases}
\end{equation}
and $\vect{\mathcal{E}}(\phi)=-\omega\d \vect{\mathcal{A}}(\phi)/\d \phi$. Here $N$ is the pulse duration in units of the laser period $2\pi/\omega$, $\phi_0$ is the CEP of the pulse and the $\sin^4$-envelope ensures that the electric field and its derivative vanish at $\phi=0$ and $\phi=2N\pi$. As we have already mentioned, the interaction of the electron and the external plane-wave field is taken into account exactly in the calculations by employing Volkov states as quantum in- and out-states of the electron and by working in the Furry picture \cite{Landau_b_4}. The interaction between the electron and the quantized photon field instead scales with the fine-structure constant $\alpha_{QED}=e^2\approx 1/137$ and in the parameter regime of our interest here it can be accounted for perturbatively up to first order. In this approximation the electromagnetic spectrum emitted by the laser-driven electron can be calculated from the probability that the electron emits one photon. We assume that the incoming electron has spin $s$ and is counter-propagating with the laser pulse, therefore its initial four-momentum is $p^{\mu}=\left(\epsilon,0,0,-p\right)$, with $\epsilon=\sqrt{m^2+p^2}$. The outgoing electron instead has spin $s'$ and four-momentum $p^{\prime\,\mu}=\left(\epsilon',\bm{p}'\right)$, with $\epsilon'=\sqrt{m^2+|\bm{p}'|^2}$. Finally, the emitted photon has four-momentum $k^{\prime\,\mu}=\left(\omega',\bm{k}'\right)$, with $\omega'=|\bm{k}'|$ and its polarization states are described by the four-vectors $\varepsilon_{r'}^{\prime\,\mu}$, with $r'\in\{1,2\}$. Then, the transition matrix element $S_{fi}$ of our process can be cast into the convenient form $S_{fi}=  (2\pi)\delta\left(p'_-+k'_--p_-\right)(2\pi)^2 \delta\left(\bm{p}'_{\perp}+\bm{k}'_{\perp}-\bm{p}_{\perp}\right)M_{fi}$, where for a general four-momentum $q^{\mu}=(q^0,\bm{q})$ the notation $q_-=q^0-q_z$ and $\bm{q}_{\perp}=(q_x,q_y)$ has been introduced. The precise expression of the amplitude $M_{fi}$ is rather involved and it is not necessary to report it here. We only note that it can be written as $M_{fi}=\sum_{j=0}^2 c_jf_j$, where $c_j$ are coefficients weakly dependent on the physical parameters of the problem and where the three functions $f_j$ with $f_j=\int_{-\infty}^{\infty}\d \eta\psi_{\mathcal{A}}^j(\eta)
e^{i\int_0^{\eta} \d\eta'[\alpha\psi_{\mathcal{A}}(\eta')+\beta\psi_{\mathcal{A}}^2(\eta')+\gamma]}$ contain all the relevant dynamical information of the process. Here we have introduced the important parameters $\alpha = -m\xi k'_x/(kp')$, $\beta = m^2\xi^2k'_-/2p_-(kp')$ and $\gamma = \omega' \left(\epsilon - p \cos\vartheta\right)/(kp')$, with $(kp')=k_{\mu}p^{\prime\,\mu}=\omega(\epsilon'-p'_3)=\omega[\epsilon+p-\omega'(1+\cos\vartheta)]$ and $\pi-\vartheta$ and $\varphi$ being the spherical angular coordinates of the emitted photon, assuming the positive $z$ axis as the polar axis. It can be shown that the function $f_0$ can be expressed in terms of the functions $f_1$ and $f_2$ as $f_0=-(\beta f_2+\alpha f_1)/\gamma$. Starting from the above quantity $M_{fi}$, one can calculate the emitted energy spectrum $\d\mathscr{E}/\d \Omega\d\omega'$ (average energy emitted between $\omega'$ and $\omega'+\d\omega'$, in the solid angle $\d\Omega=\sin\vartheta \d\vartheta\d\varphi$) as
\begin{equation}
\label{ergspec}
\frac{\d\mathscr{E}}{\d \Omega\d\omega'} = \frac{\omega^{\prime\, 3}}{16\pi^3}\sum_{s,s',r'}\left|M_{fi}\right|^2.
\end{equation}
It can be seen that, as it must be, the above expression of the emitted energy spectrum reduces to its classical counterpart when the energy of the emitted photon is much smaller than the initial electron energy, i. e. in the limit $\omega'\ll \epsilon$. In turn, this occurs if the parameter $\chi=(\epsilon+p)\xi \omega/m^2$ is much smaller than unity \cite{Ritus_Review}. From a physical point of view, the parameter $\chi$ is the laser's electric field amplitude in units of the QED critical field $E_{cr}=m^2/|e|$ in the rest frame of the incoming electron. Effects of the laser's pulse shape on \emph{classical} multiphoton Thomson scattering ($\chi\ll 1$) at laser intensities around the relativistic threshold have been studied in \cite{Lan_2007,Boca_2009,Harvey_2009}. CEP effects have also been investigated in Schwinger electron-positron pair production in time-dependent electric fields \cite{Hebenstreit_2009}, but, for pair production to occur at all in that case, laser intensities have been considered larger than $10^{27}\;\text{W/cm$^2$}$.

Now, as we have mentioned, we are interested in the ultra-relativistic regime in which $\xi\gg 1$. Also, we will consider situations in which $m\xi\sim \epsilon$, where the interplay between the initial electron energy and the laser intensity produces rich dynamics of the electron in the laser field. In this parameter regime, if $\chi\ll 1$ (classical case) the electron mainly emits frequencies of the order of $\omega'\sim\omega\xi^3$, while if $\chi\gtrsim 1$ (quantum case) the electron mainly emits in the energy range $\omega'\sim \epsilon$. In both cases, one can see that the three parameters $\alpha$, $\beta$ and $\gamma$ appearing in the exponential in the functions $f_j$ are all of the same order and very large. This implies that the functions $f_j$ can be evaluated by applying the saddle-point method to the integrals in $\eta$. In the classical limit, this circumstance reflects the following physical feature: the spectrum emitted by an ultra-relativistic electron along a direction $(\vartheta,\varphi)$ is mainly determined by those parts of the electron trajectory where its velocity points along $(\vartheta,\varphi)$, within a small angle of the order of $m/\epsilon_e\ll 1$, with $\epsilon_e$ being the electron's energy at the emission time, and the other parts of the trajectory give an exponentially small contribution \cite{Landau_b_2_1975}. This is the physical reason why the energy spectrum emitted by an ultra-relativistic electron provides detailed information about the electron trajectory (in the classical regime) and, in turn, about the precise form of the driving external field. As we will show below, this last feature remains true also in the quantum regime. In fact, from the general theory of the saddle-point method it follows in our case that in the region of parameters in which the saddle points have a large imaginary part, the functions $f_j$ are exponentially smaller than in those regions in which the saddle points are almost real \cite{Ritus_Review}. If we fix the initial energy of the electron, the laser intensity and the energy of the emitted photon, the functions $f_j$ depend only on the photon emission angles $\vartheta$ and $\varphi$. We have shown that the region where the saddle points are almost real (their imaginary part scales as $1/\xi\ll 1$) corresponds in the classical limit exactly to the angular region in $\vartheta$ and $\varphi$, which is spanned by the electron velocity when it moves inside the laser pulse. This is fully consistent with the mentioned classical circumstance that an ultra-relativistic electron emits almost only along its velocity. By consequently imposing that the saddle points $\eta_s$ of the phase in $f_j$ are real, the following condition must be fulfilled
\begin{equation}
\psi_{\mathcal{A},\text{min}}\leq \frac{\epsilon + p}{m\xi}\tan\left(\frac{\vartheta}{2}\right) \cos\varphi \leq \psi_{\mathcal{A},\text{max}}, \label{angcondition}
\end{equation}
where $\psi_{\mathcal{A},\text{min}}$ ($\psi_{\mathcal{A},\text{max}}$) denotes the minimal (maximal) value of the function $\psi_{\mathcal{A}}(\phi)=\int_0^{\phi}\d\phi'\psi_{\mathcal{E}}(\phi')$. The values of $\psi_{\mathcal{A},\text{min}}$ and $\psi_{\mathcal{A},\text{max}}$ depend on the CEP $\phi_0$ and the above equation yields the bounding angles for the emission cone as functions of the CEP. It is worth observing that, since we work in the ultra-relativistic regime $\epsilon\gg m$, then $p\approx\epsilon$ and Eq. (\ref{angcondition}) provides universal conditions depending (apart from the shape of the laser field, of course) only on the ratio $\epsilon/m\xi$.

The above considerations equally apply in the classical and in the quantum regime and the spectrum (\ref{ergspec}) in the limit $\chi\ll 1$ goes to the classical expression derived by first solving the Lorentz equation and then plugging the resulting trajectory into the Li\'{e}nard-Wiechert potentials (see Eq. (66.9) in \cite{Landau_b_2_1975}). However, the classical formulas hold in principle for an arbitrary external field. Moreover, it can be shown that the above equation (\ref{angcondition}) can also be employed for an external potential of the form $\vect{\mathcal{A}}(\phi;x,y,z)=\mathcal{A}\psi_{\mathcal{A}}(\phi;x,y,z)\,\hat{\vect{x}}$  slowly-varying with respect to $x$, $y$ and $z$, by simply substituting the function $\psi_{\mathcal{A}}(\phi)$ with $\psi_{\mathcal{A}}(\phi;x,y,z)$. In Fig. 1 we show the dependence of the angular emission region in $\vartheta$  as a function of the CEP, obtained from Eq. (\ref{angcondition}) in the two cases $m\xi = 2 \epsilon$ (Fig. 1a) and $m\xi=\epsilon/7.5$ (Fig. 1b). In both cases $N=2$. In view of the first numerical example worked out below, in Fig. 1a we consider a Gaussian beam with carrier wavelength $\lambda=1.2\;\text{$\mu$m}$ ($\omega=1\;\text{eV}$), spot radius $w=2\;\text{$\mu$m}$ at zero order in the ratio $w/z_r<1$, with $z_r=\omega w^2/2$ being the laser's Rayleigh length \cite{Salamin_2002} (we will see that our formula (\ref{angcondition}) with $\psi_{\mathcal{A}}(\phi)\to\psi_{\mathcal{A}}(\phi;x,y,z)$ works well also at such tight focusing). In Fig. 1b the case of an external plane wave is considered. The radiation is confined around the azimuthal angles $\varphi=0$ and $\varphi=\pi$ within a small aperture angle of roughly $\Delta\varphi\sim m/\epsilon\ll 1$, therefore we consider only the directions exactly at $\varphi=0$ and at $\varphi=\pi$, corresponding to $\vartheta\ge 0$ and $\vartheta<0$ in Fig. 1, respectively.
\begin{figure}[ht]
\centering
\includegraphics[width=0.9\columnwidth]{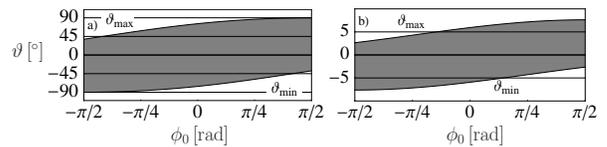}
\caption{Analytical angular emission range as a function of the CEP obtained from Eq. (\ref{angcondition}) for a two-cycle pulse with $m\xi = 2 \epsilon$ (part a)) and $m\xi=\epsilon/7.5$ (part b)). In part a) the external field is a Gaussian beam with carrier angular frequency $\omega=1\;\text{eV}$ and spot radius $w=2\;\text{$\mu$m}$. In part b) the external field is a plane wave.}
\end{figure}
The figure shows a one-to-one dependence of the emission range $\Delta\vartheta=\vartheta_{\text{max}}-\vartheta_{\text{min}}$ on the CEP $\phi_0$ ($\vartheta_{\text{min}}$ and $\vartheta_{\text{max}}$ are the bounding angles obtained from Eq. (\ref{angcondition})). From Eq. (\ref{angcondition}) one also sees that the most convenient range of parameters in terms of accuracy in the determination of the CEP is at $m\xi\sim \epsilon$ (if $m\xi\gg\epsilon$, then the electron emits almost exclusively into a cone along the laser propagation direction with an angular aperture of the order of $1/\xi$ independently of the CEP).

In order to show quantitatively the features of our method we consider below two examples. In the first example, we aim to investigate a rather realistic situation in which the external field is modeled as a focused Gaussian beam also including longitudinal field components up to first order in $w/z_r$ \cite{Salamin_2002}. By recalling the laser's parameters envisaged at the PFS \cite{PFS}, we set $\omega=1\;\text{eV}$, $N=2$ (corresponding to $\tau=8\;\text{fs}$), $w=2\;\text{$\mu$m}$ and $\xi=100$ ($I=10^{22}\;\text{W/cm$^2$}$). Moreover, it can be seen that for the electron densities of beams obtained via laser wakefield acceleration \cite{Dream_beams}, the relevant (high-energy) part of the spectrum results essentially from incoherent emission. We consider an electron beam with a three-dimensional Gaussian spatial distribution with waists $w_{e,x}=w_{e,y}=5\;\text{$\mu$m}$ and $w_{e,z}=8\;\text{$\mu$m}$ and with a Gaussian energy distribution with central energy $\epsilon=26\;\text{MeV}$ (such that $m\xi\approx 2\epsilon$ like in Fig. 1a) and waist $w_{e,\epsilon}$ such that $w_{e,\epsilon}/\epsilon=2\;\%$ \cite{Dream_beams}. Therefore, since $\chi\approx 2\times 10^{-2}$, we can calculate the spectrum by employing the classical formula valid for a single electron \cite{Landau_b_2_1975} and then averaging it over the electron distribution (in the numerical example we have considered a beam with $N_e=300$ electrons and ensured that our results are not significantly altered by increasing $N_e$). In this example (see Figs. 2a and 2b) we aim to show the sensitivity of our method and we show two energy spectra for the two different CEPs $\phi_0 = -\pi/10$ (Fig. 2a) and $\phi_0 = -\pi/5$ (Fig. 2b).
\begin{figure}[ht]
\centering
\includegraphics[width=\linewidth]{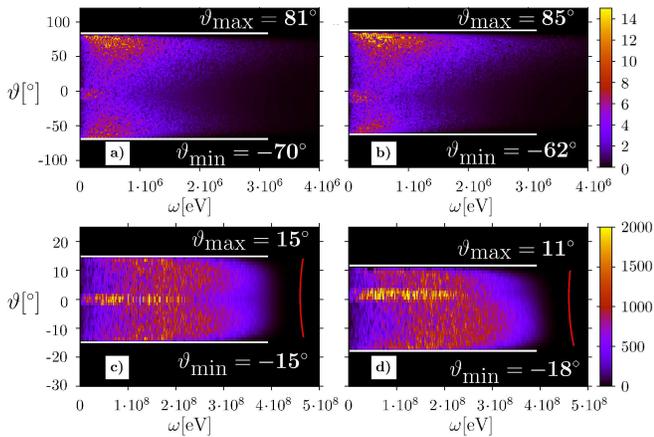}
\caption{(Color online) Energy emission spectra $\d\mathscr{E}/\d \Omega\d\omega'$ in $\text{sr$^{-1}$}$ via Eq. (66.9) in \cite{Landau_b_2_1975} (parts a) and b)) and via Eq. (\ref{ergspec}) (parts c) and d)) for the two sets of parameters described in the text. In parts a) and b) quantum effects are negligible and it is $\phi_0 = -\pi/10$ (part a)) and $\phi_0 = -\pi/5$ (part b)). In part c) and d) quantum effects are important and it is $\phi_0 = 0$ (part c)) and $\phi_0 = \pi/4$ (part d)) (the almost vertical red line  indicates here the quantum cut-off frequency $\omega'_M=(\epsilon+p)/(1+\cos\vartheta)$). The horizontal white lines indicate the boundary of the emission range determined analytically from Eq. (\ref{angcondition}) generalized to the case of a Gaussian beam for parts a) and b).}
\label{emission1}
\end{figure}
The white horizontal lines indicate the values of $\vartheta_{\text{min}}$ and $\vartheta_{\text{max}}$ as predicted by Eq. (\ref{angcondition}) generalized to the case of a Gaussian beam. The plots show the very good agreement between the analytical predictions and the numerical results. Moreover, it is evident that the angular aperture of the emission region is very sensitive to the CEP: a small change $|\Delta\phi_0|=\pi/10$ in the CEP changes the minimum (maximum) emission angle by about $3^{\circ}$ ($5^{\circ}$). In Figs. 2c and 2d we consider another example, in which an ultra-strong attosecond XUV pulse, as those theoretically envisaged in \cite{Tsakiris_2006} is employed ($\omega=50\;\text{eV}$, $\tau=160\;\text{as}$ and $\xi=20$, which can be obtained if the field is focused to about $w=100\;\text{nm}$) and an electron with initial energy of $75\;\text{MeV}$ (corresponding to $7.5\,m\xi$ as in Fig. 1b). In this case, $\chi=0.6$ and quantum effects are already important. Since here $\lambda=25\;\text{nm}$ and $w=100\;\text{nm}$, the quantum results valid in the plane-wave approximation should apply with sufficient accuracy. With these parameters the method turns to be less precise as compared to the above example and a change in the CEP of about $\pi/4$ produces a change in the angular aperture of a few degrees (in Fig. 2c it is $\phi_0=0$ while in Fig. 2d $\phi_0=\pi/4$). In this case we observe an excellent agreement between the numerical and the analytical values of $\vartheta_{\text{min}}$ and $\vartheta_{\text{max}}$. Moreover, the almost vertical red line shows the position of the cut-off emission frequency $\omega'_M$ determined from the analytical formula $\omega'_M=(\epsilon+p)/(1+\cos\vartheta)$ \cite{Ritus_Review}. This is a typical quantum effect: due to energy-momentum conservation, the electron cannot emit a photon at an angle $\vartheta$ with energy larger than or equal to $\omega'_M$. Therefore, also the quantum cut-off frequency $\omega'_M$ is affected by the CEP, through the emission aperture $\Delta\vartheta$. Finally, we note that the bright features of the spectra in Figs. 2c and 2d along the direction $\vartheta=0$ can be explained classically: in a relatively large part of the trajectory inside the laser field the electron velocity is observed to point just along that direction.

Experimental uncertainties in the laser intensity may alter our results. The intensity of a strong optical laser beam can be measured nowadays with a relative uncertainty $\Delta I\,/I$ of about $10\;\text{\%}$ \cite{Emax}. If one includes a corresponding uncertainty $\Delta\xi$ in the value of $\xi$ in Eq. (\ref{angcondition}), one obtains that the induced uncertainty $\Delta\vartheta_{\text{min/max}}$ in the predicted values $\vartheta_{\text{min/max}}$ is approximately given by $\Delta\vartheta_{\text{min/max}}\approx 4m\epsilon|\psi_{\mathcal{A},\text{min/max}}|\Delta \xi/(4\epsilon^2+m^2\xi^2|\psi_{\mathcal{A},\text{min/max}}|^2)$. In the example in Figs. 2a and 2b we obtain $\Delta\vartheta_{\text{min/max}}\approx 2.8^{\circ}$. In the second example, it is difficult to estimate the value of $\Delta I\,/I$. However, since $\epsilon\approx 7.5\,m\xi$ then $\Delta\vartheta_{\text{min/max}}\approx (\Delta\xi/\xi)/7.5$ and even an uncertainty of about $50\;\%$ in the intensity is acceptable. Thus in both cases we can conclude that these uncertainties do not conceal the effect of the CEP. By repeating the simulation with a different temporal pulse shape (a Gaussian one) we have observed alterations to the spectra, much smaller than those due to the uncertainty in the laser intensity. We finally note that for pulses comprising more than three laser cycles the discussed effect is too small to be significant. On the other hand, the CEP effect is even larger than here for pulses including only one laser cycle.


\begin{thebibliography}{39}
\expandafter\ifx\csname natexlab\endcsname\relax\def\natexlab#1{#1}\fi
\expandafter\ifx\csname bibnamefont\endcsname\relax
  \def\bibnamefont#1{#1}\fi
\expandafter\ifx\csname bibfnamefont\endcsname\relax
  \def\bibfnamefont#1{#1}\fi
\expandafter\ifx\csname citenamefont\endcsname\relax
  \def\citenamefont#1{#1}\fi
\expandafter\ifx\csname url\endcsname\relax
  \def\url#1{\texttt{#1}}\fi
\expandafter\ifx\csname urlprefix\endcsname\relax\def\urlprefix{URL }\fi
\providecommand{\bibinfo}[2]{#2}
\providecommand{\eprint}[2][]{\url{#2}}

\bibitem{Bonvalet} 
  \bibinfo{author}{A. Bonvalet et al.},
  \bibinfo{journal}{Appl. Phys. Lett.} \textbf{\bibinfo{volume}{67}},
  \bibinfo{pages}{2907} (\bibinfo{year}{1995}).
\bibitem{Krauss_2010} G. Krauss et al., Nature Photonics \textbf{4}, 33 (2010).
\bibitem{Cavalieri_2007} A.L. Cavalieri et al., New J. Phys. \textbf{9}, 242 (2007).

\bibitem{Goulielmakis_2008} E. Goulielmakis et al, Science \textbf{320}, 1614 (2008).

\bibitem{Krausz_2009} F. Krausz and M. Ivanov, Rev. Mod. Phys. \textbf{81}, 163 (2009).

\bibitem{Paulus_2003} G. G. Paulus et al., Nature \textbf{414}, 182 (2001);
  \bibinfo{author}{\bibnamefont{G.~G.} \bibnamefont{Paulus}},
  \bibinfo{journal}{Phys. Rev. Lett.} \textbf{\bibinfo{volume}{91}},
  \bibinfo{eid}{253004} (\bibinfo{year}{2003}).
\bibitem{Wittmann_2009} T. Wittmann et al., Nature Phys. \textbf{5}, 357 (2009).
\bibitem{Goulielmakis_2004} E. Goulielmakis et al., Science \textbf{305}, 1267 (2004).
\bibitem{Kress_2006} M.Kre{\ss} et al., Nature Phys. \textbf{2}, 327 (2006).
\bibitem{Landau_b_2_1975} L. D. Landau, and E. M. Lifshitz, \textit{The Classical Theory of Fields}, Elsevier, Oxford (1975), \S~47, 66 and 73-74.
\bibitem{Tsung_2002} F. S. Tsung et al., Proc. Nat. Acad. Sc. USA \textbf{99}, 29 (2002).
\bibitem{Tsakiris_2006} G. D. Tsakiris et al., New. J. Phys. \textbf{8}, 19 (2006).
\bibitem{PFS} http://www.attoworld.de/.
\bibitem{ELI} http://www.extreme-light-infrastructure.eu/.
\bibitem{Landau_b_4} V. B. Berstetskii, E. M. Lifshitz, and L. P. Pitaevskii, \textit{Quantum Electrodynamics}
(Elsevier, Oxford, 1982), \S~40 and 101.

\bibitem{Brabec_2000} T. Brabec and F. Krausz. Rev. Mod. Phys. \textbf{72}, 545 (2000).
\bibitem{Ritus_Review} V. I. Ritus, J. Sov. Laser Res. \textbf{6}, 497 (1985).
\bibitem{Lan_2007} P. Lan et al., J. Phys. B: At. Mol. Opt. Phys. \textbf{40}, 403 (2007).
\bibitem{Boca_2009} M. Boca and V. Florescu, Phys. Rev. A \textbf{80}, 053403 (2009).

\bibitem{Harvey_2009} C. Harvey, T. Heinzl and A. Ilderton, Phys. Rev. A \textbf{79}, 063407 (2009); T. Heinzl, D. Seipt and B. K\"{a}mpfer, Phys. Rev. A \textbf{81}, 022125 (2010).
\bibitem{Hebenstreit_2009} F. Hebenstreit et al., Phys. Rev. Lett. \textbf{102}, 150404 (2009).

\bibitem{Emax} V. Yanovsky et al., Opt. Express {\bf 16}, 2109 (2008).
\bibitem{Salamin_2002} Y. I. Salamin and C. H. Keitel, Phys. Rev. Lett. \textbf{88}, 095005 (2002).

\bibitem{Dream_beams} W. P. Leemans et al., Nature Phys. \textbf{2}, 696 (2006); S. Kneip et al., Phys. Rev. Lett. \textbf{103}, 035002 (2009) and references therein.

\end{thebibliography}
\end{document}